\documentclass[nohyperref]{article}

\usepackage{microtype}
\usepackage{graphicx}
\usepackage{subfigure}
\usepackage{booktabs} 

\usepackage{hyperref}

\usepackage[accepted]{icml2022}

\usepackage{amsmath}
\usepackage{amssymb}
\usepackage{mathtools}
\usepackage{amsthm}

\usepackage[capitalize,noabbrev]{cleveref}

\theoremstyle{plain}

\theoremstyle{definition}

\theoremstyle{remark}

\usepackage[textsize=tiny]{todonotes}

\usepackage{multirow}
\usepackage{xspace}

\newcommand{\pt}{\ensuremath{p_{\text{T}}}\xspace}
\newcommand{\jetclass}{{\textsc{JetClass}}\xspace}
\newcommand{\rej}[1]{\ensuremath{\text{Rej}_{#1}}\xspace}
\newcommand{\sst}[1]{\ensuremath{{}_{\text{#1}}}\xspace}

\newcommand{\hbb}{\ensuremath{H\to b \bar{b}}\xspace}
\newcommand{\hcc}{\ensuremath{H\to c \bar{c}}\xspace}
\newcommand{\hgg}{\ensuremath{H\to g g}\xspace}

\newcommand{\hqqqq}{\ensuremath{H\to 4 q}\xspace}
\newcommand{\hlvqq}{\ensuremath{H\to \ell \nu q q'}\xspace}
\newcommand{\tbqq}{\ensuremath{t\to b q q'}\xspace}
\newcommand{\tblv}{\ensuremath{t\to b \ell \nu}\xspace}
\newcommand{\wqq}{\ensuremath{W\to q q'}\xspace}
\newcommand{\zqq}{\ensuremath{Z\to q \bar{q}}\xspace}
\newcommand{\qgj}{\ensuremath{q/g}\xspace}

\newcommand{\MADGRAPH}{\textsc{MadGraph}\xspace}
\newcommand{\MCATNLO} {\textsc{mc@nlo}\xspace}
\newcommand{\MGvATNLO}{\MADGRAPH{}5\_a\MCATNLO}
\newcommand{\PYTHIA} {{\textsc{pythia}}\xspace}
\newcommand{\DELPHES} {{\textsc{Delphes}}\xspace}

\begin{document}

\twocolumn[
\icmltitle{Particle Transformer for Jet Tagging}



\icmlsetsymbol{equal}{*}

\begin{icmlauthorlist}
\icmlauthor{Huilin Qu}{cern}
\icmlauthor{Congqiao Li}{pku}
\icmlauthor{Sitian Qian}{pku}
\end{icmlauthorlist}

\icmlaffiliation{cern}{CERN, Geneva, Switzerland}
\icmlaffiliation{pku}{School of Physics, Peking University, Beijing, China}

\icmlcorrespondingauthor{Huilin Qu}{huilin.qu@cern.ch}
\icmlcorrespondingauthor{Congqiao Li}{licongqiao@pku.edu.cn}
\icmlcorrespondingauthor{Sitian Qian}{stqian@pku.edu.cn}

\icmlkeywords{Transformer, Jet Tagging, Dataset, ICML}

\vskip 0.3in
]



\printAffiliationsAndNotice{}  

\begin{abstract}
Jet tagging is a critical yet challenging classification task in particle physics. 
While deep learning has transformed jet tagging and significantly improved performance, the lack of a large-scale public dataset impedes further enhancement. 
In this work, we present \jetclass, a new comprehensive dataset for jet tagging. 
The \jetclass dataset consists of 100\,M jets, about two orders of magnitude larger than existing public datasets. 
A total of 10 types of jets are simulated, including several types unexplored for tagging so far. 
Based on the large dataset, we propose a new Transformer-based architecture for jet tagging, called Particle Transformer (ParT). 
By incorporating pairwise particle interactions in the attention mechanism, ParT achieves higher tagging performance than a plain Transformer and surpasses the previous state-of-the-art, ParticleNet, by a large margin. 
The pre-trained ParT models, once fine-tuned, also substantially enhance the performance on two widely adopted jet tagging benchmarks.
The dataset, code and models are publicly available at~\url{https://github.com/jet-universe/particle_transformer}.
\end{abstract}

\section{Introduction}
\label{sec:intro}

Machine learning has revolutionized how large-scale data samples are analyzed in particle physics and greatly increased the discovery potential for new fundamental laws of nature \cite{Radovic:2018dip}. Specifically, deep learning has transformed how \textit{jet tagging}, a critical classification task at high-energy particle colliders such as the CERN LHC, is performed, leading to a drastic improvement in its performance \cite{Kogler:2018hem,Larkoski:2017jix}.

\begin{figure}[tb]
\begin{center}
\centerline{\includegraphics[width=\columnwidth]{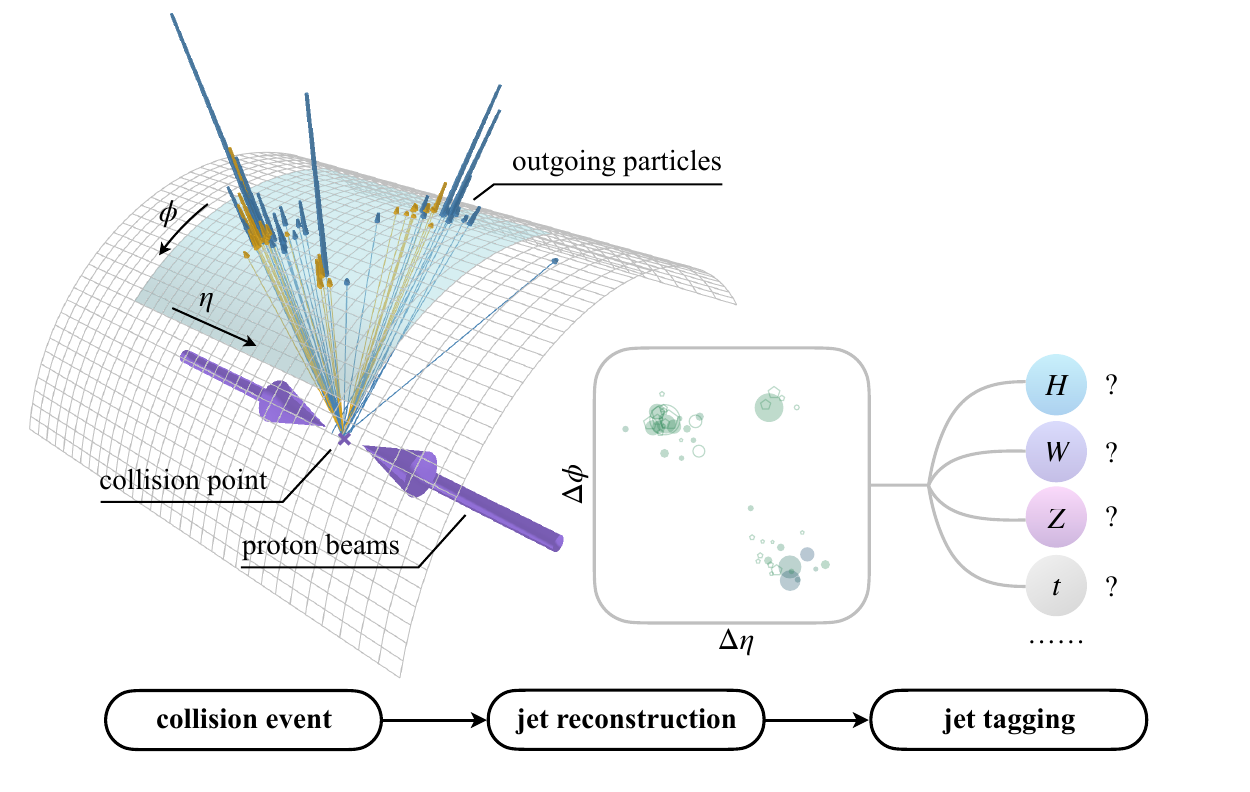}}
\caption{Illustration of jet tagging at the CERN LHC. High-energy proton-proton collisions at the LHC can produce new unstable particles that decay and yield a collimated spray of outgoing particles. These outgoing particles are measured by complex particle detector systems, and jets can be built (``reconstructed'') from these measured particles. The goal of jet tagging is to classify the jets and identify those arising from particles of high interest, e.g., the Higgs boson, the $W$ or $Z$ boson, or the top quark.}
\label{fig:jet-tagging}
\end{center}
\vskip -0.4in
\end{figure}

At the CERN LHC, two beams of protons are accelerated to nearly the speed of light and made to collide at a frequency of 40 million times per second (40\,MHz). Such high-energy collisions can create new unstable particles, which then decay and produce sprays of outgoing particles. Complex detector systems, such as the general-purpose ATLAS \cite{ATLAS:2008xda} and CMS \cite{CMS:2008xjf} detectors with $\mathcal{O}$(100\,M) individual sensors of various types, are used to measure the positions, trajectories, energies, and momenta of the outgoing particles. From these measurements, an \textit{event} is reconstructed for each collision. The primary goal in the analysis of the collision data is to identify events involving novel physics processes, an example of which is the discovery of the Higgs boson \cite{ATLAS:2012yve,CMS:2012qbp}.

\begin{figure*}[t]
\begin{center}
\centerline{\includegraphics[width=\textwidth]{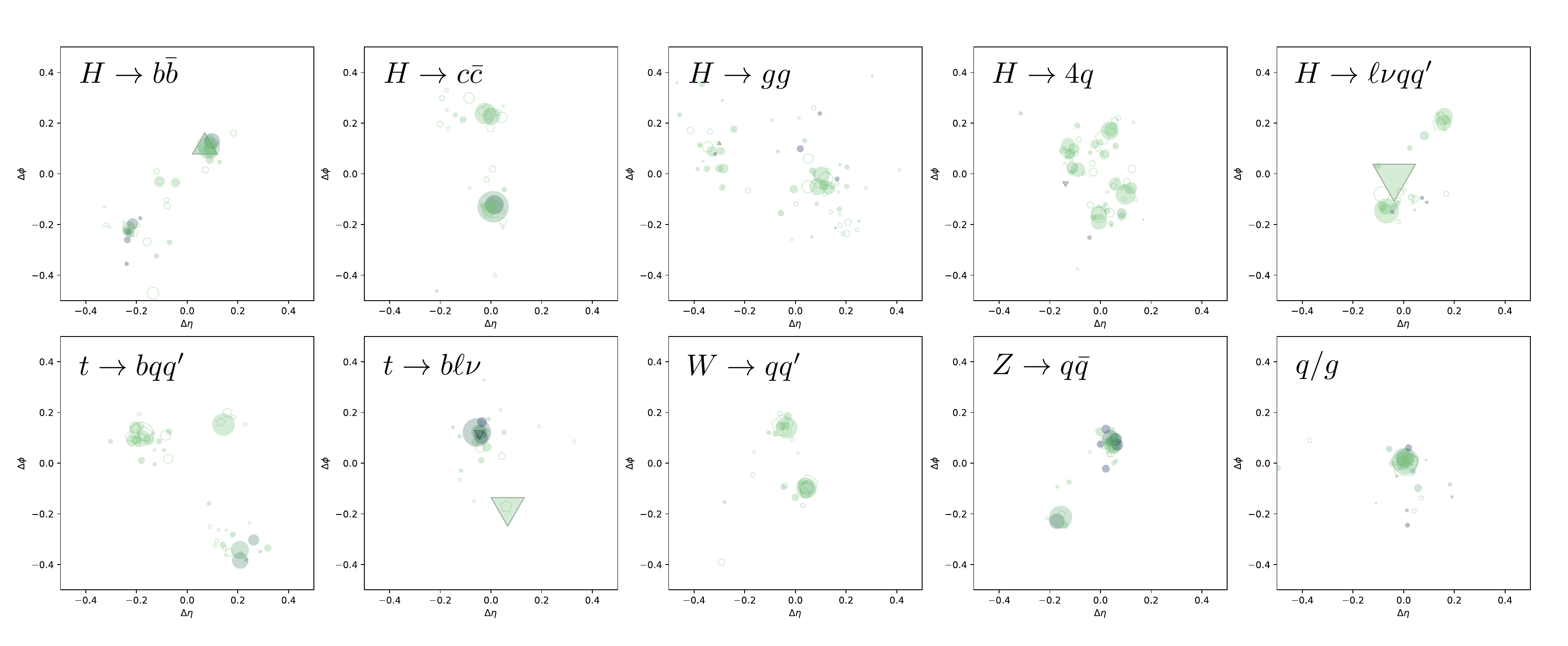}}
\vskip -0.2in
\caption{Examples of the 10 types of jets in the \jetclass dataset, viewed as particle clouds. Each particle is displayed as a marker, with its coordinates corresponding to the flying direction of the particle, and its size proportional to the energy. The circles, triangles (upward- or downward-directed), and pentagons represent the particle types, which are hadrons, leptons (electrons or muons), and photons, respectively. The solid (hollow) markers stand for electrically charged (neutral) particles. The marker color reflects the displacement of the particle trajectory from the interaction point of the proton-proton collision, where a larger displacement results in more blue.}
\label{fig:dataset}
\end{center}
\vskip -0.4in
\end{figure*}

A crucial step in the data analysis process is jet tagging. A \textit{jet} refers to a collimated spray of outgoing particles. Jet tagging is the process of identifying the type of particle that initiates a jet. It is essentially a classification task that aims to distinguish jets arising from particles of interest, such as the Higgs boson or the top quark, from other less interesting types of jets. Jet tagging is a challenging task because the particle initiating a jet can radiate, and the radiated particles further produce more particles, leading to a cascade of $\mathcal{O}(10)$ to $\mathcal{O}(100)$ particles at the end. The radiation also smears the characteristics of the initial particle and makes the identification very difficult. 

Traditional approaches for jet tagging rely on hand-crafted features motivated by the principles of quantum chromodynamics (QCD), the theory governing the evolution of particles inside a jet. The rise of deep learning has led to a plethora of new approaches \cite{Larkoski:2017jix}. The prevailing ones represent a jet as a \textit{particle cloud}, i.e., an unordered and variable-sized set of the outgoing particles, as illustrated in \cref{fig:jet-tagging}. Based on the particle cloud representation, ParticleNet \cite{Qu:2019gqs} adapts the Dynamic Graph CNN architecture \cite{dgcnn} and achieves substantial performance improvement on two representative jet tagging benchmarks. 
Since then, several new models (e.g., \citet{Mikuni:2020wpr,Mikuni:2021pou,Shimmin:2021pkm}) have been proposed, but no significant performance improvement has been reported so far. We deem the lack of a sufficiently large public dataset an impeding factor.

In this work, we advocate for \jetclass, a new large and comprehensive dataset to advance deep learning for jet tagging. The \jetclass dataset \cite{jetclass} consists of 100\,M jets for training, about two orders of magnitude larger than existing public datasets. It also includes more types of jets, several of which have not been explored for tagging yet but are promising for future applications at the LHC. 

Based on this dataset, we propose Particle Transformer (ParT), a new Transformer-based architecture for jet tagging. We demonstrate that Transformer architectures, together with a large dataset, can reach powerful performance on jet tagging. We introduce a small modification to the attention mechanism by incorporating a new term characterizing pairwise particle interactions. The resulting ParT achieves significantly higher performance than a plain Transformer and surpasses the previous state-of-the-art, ParticleNet, by a large margin. We also apply the pre-trained ParT models to two widely adopted jet tagging benchmarks with fine-tuning and observe a substantial gain on these tasks.

\section{The \textsc{JetClass} Dataset}
\label{sec:dataset}

We provide an overview of the new \jetclass dataset in this section. The dataset includes a total of 10 types of jets. 
Representative jets of each type are visualized as particle clouds in \cref{fig:dataset}.
The jets in this dataset generally fall into two categories. 
The \textit{background} jets are initiated by light quarks or gluons (\qgj) and are ubiquitously produced at the LHC. 
The \textit{signal} jets are those arising either from the top quarks~($t$), or from the $W$, $Z$ or Higgs~($H$) bosons. For top quarks and Higgs bosons, we further consider their different decay modes as separate types, because the resulting jets have rather distinct characteristics and are often tagged individually. 
The use of jet tagging typically involves selecting one (or a few) specific type of signal jets with high confidence, and rejecting background jets as much as possible, since the background jets usually appear orders of magnitude more frequently than the targeted signal jets. 
Note that for several types of signal jets in this dataset, such as \hqqqq, \hlvqq, and \tblv, no dedicated methods have been developed so far to tag them. However, as we will demonstrate in \cref{sec:exp-jet-class}, these types of jets can also be cleanly tagged with deep learning approaches, opening up new possible territories for jet tagging at the LHC.

\textbf{Simulation setup.} Jets in this dataset are simulated with standard Monte Carlo event generators used by LHC experiments. The production and decay of the top quarks and the $W$, $Z$ and Higgs bosons are generated with {\MGvATNLO} \cite{Alwall:2014hca}. We use \PYTHIA \cite{Sjostrand:2014zea} to evolve the produced particles, i.e., performing parton showering and hadronization, and produce the final outgoing particles\footnote{We include multiple parton interactions but omit pileup interactions in the simulation.}.
To be close to realistic jets reconstructed at the ATLAS or CMS experiment, detector effects are simulated with \DELPHES \cite{deFavereau:2013fsa} using the CMS detector configuration provided in \DELPHES. In addition, the impact parameters of electrically charged particles are smeared to match the resolution of the CMS tracking detector \cite{CMS:2014pgm}. Jets are clustered from \DELPHES E-Flow objects with the anti-$k_{\text{T}}$ algorithm \cite{Cacciari:2008gp, Cacciari:2011ma} using a distance parameter $R=0.8$. Only jets with transverse momentum in 500--1000\,GeV and pseudorapidity $|\eta|<2$ are considered. For signal jets, only the ``high-quality'' ones that fully contain the decay products of initial particles are included\footnote{We require all the quarks ($q$) and charged leptons (electrons or muons, denoted $\ell$) from the decay of the top quark or the $W$, $Z$ or Higgs boson satisfy $\Delta R(\text{jet}, q/\ell)<0.8$, where $\Delta R(a, b)\equiv\sqrt{(\eta_{a} - \eta_{b})^2 + (\phi_{a} - \phi_{b})^2}$, in which $\eta$ ($\phi$) is the pseudorapidity (azimuthal angle) of the momentum of the jet or the particle.}.

\textbf{Input features.} The dataset provides all constituent particles of each jet as inputs for jet tagging. Note that the number of particles varies from jet to jet, typically between 10 and 100, with an average of 30--50 depending on the jet type. For each particle of a jet, three categories of features are provided:
\begin{itemize}
    \vspace{-.2cm}
    \item \textbf{Kinematics.} This includes the energy and momentum, described by the 4-vector $(E, p_{x}, p_{y}, p_{z})$ in units of GeV, which are the most fundamental quantities measured by a particle detector. All other kinematic variables can be computed from the 4-vectors.
    \vspace{-.2cm}
    \item \textbf{Particle identification.} This includes the electric charge, with values of $\pm1$ (positively/negatively charged particles) and 0 (neural particles), and the particle identity determined by the detector systems. For the latter, a 5-class one-hot encoding is used to be consistent with current LHC experiments: charged hadron ($\pm211,\pm321,\pm2212$), neutral hadron (0), electron ($\pm11$), muon ($\pm13$), and photon (22). The particle identification information is especially important for tagging jets involving a charged lepton, e.g., \hlvqq and \tblv, as leptons can be almost unambiguously identified at the LHC.
    \vspace{-.2cm}
    \item \textbf{Trajectory displacement.} This includes the measured values and errors of the transverse and longitudinal impact parameters of the particle trajectories in units of mm, in total 4 variables. These measurements are only available for electrically charged particles, and a value of 0 is used for neutral particles. The trajectory displacement information is critical for tagging jets involving a bottom ($b$) or charm ($c$) quark \cite{CMS:2020poo}, such as \hbb, \hcc, \tbqq, etc., but is missing from most of the existing datasets. 
    \vspace{-.2cm}
\end{itemize}

\textbf{Training, validation and test sets.} The training set consists of 100\,M jets in total, equally distributed in the 10 classes. An additional set of 500\,k jets per class (in total 5\,M) is intended for model validation. For the evaluation of performance, a separate test set with 2\,M jets in each class (in total 20\,M) is provided. 

\textbf{Evaluation metrics.} To thoroughly evaluate the performance of deep learning models on this dataset, we advocate for a series of metrics. Since jet tagging on this dataset is naturally framed as a multi-class classification task, two common metrics, i.e., the accuracy and the area under the ROC curve (AUC)\footnote{The AUC can be calculated using \texttt{roc\_auc\_score} in scikit-learn with \texttt{average='macro'} and \texttt{multi\_class='ovo'}.} are adopted to quantify the overall performance. In addition, we propose the \textit{background rejection} (i.e., the inverse of the false positive rate) at a certain signal efficiency (i.e., the true positive rate, TPR) of $X$\%, i.e., 
\begin{equation}
    \rej{X\%}\equiv 1/\text{FPR} \text{ at TPR}=X\%,    
\end{equation}
for each type of signal jets. By default, the \qgj jets are considered as the background, as is the case in most LHC data analyses, and each of the other 9 types of jets can be considered as the signal. The signal efficiency (TPR) for each signal type is chosen to be representative of actual usages at the LHC experiments and is typically 50\%. It is increased to 99\% (99.5\%) for \hlvqq (\tblv), as these types of jets have more distinct characteristics and can be more easily separated from \qgj jets. Since the definition of the \rej{X} metric involves only two classes, i.e., the signal class under consideration ($S$) and the background class ($B$), the TPR and FPR are evaluated using a two-class score, 
\begin{equation}
    \text{score}_{S \text{vs} B} \equiv \frac{\text{score}(S)}{\text{score}(S)+\text{score}(B)},    
\end{equation}
where $\text{score}(S)$ and $\text{score}(B)$ are the softmax outputs for class $S$ and $B$, respectively, to achieve optimal performance for $S$ vs $B$ separation. 
This is aligned with the convention adopted by the CMS experiment \cite{CMS:2020poo}. Note that the background rejection metric, although rarely used in vision or language tasks, is actually a standard metric for jet tagging because it is directly related to the discovery potential at the LHC experiments. A factor of two increase in background rejection can lead to about 40\% increase in the discovery potential, which would otherwise require a dataset of twice the size, or in other words, doubling the running time of the LHC.

\section{Related Work}
\label{sec:related-work}

\textbf{Jet tagging with deep learning.}
Deep learning approaches have been proposed extensively to improve jet tagging. Previous models handle jets with different representations, e.g., images~\cite{deOliveira:2015xxd}, sequences~\cite{Guest:2016iqz}, trees~\cite{Louppe:2017ipp}, graphs~\cite{henrionneural}, with corresponding deep learning architectures such as 2D CNNs, recurrent or recursive networks, and graph neural networks. More recently, the particle cloud representation~\cite{Komiske:2018cqr,Qu:2019gqs}, analogous to point clouds, which treats a jet as a permutation-invariant set of particles as visualized in \cref{fig:dataset}, has been proposed. The Deep Sets \cite{NIPS2017_f22e4747} and Dynamic Graph CNN \cite{dgcnn} architectures are adapted for jet tagging, resulting in the Energy Flow Network \cite{Komiske:2018cqr} and the state-of-the-art, ParticleNet \cite{Qu:2019gqs}, respectively. Since then, particle clouds have become the prevailing representation of jets and more architectures based on GAPNet \cite{chen2019gapnet,Mikuni:2020wpr}, the Point Cloud Transformer \cite{guoPCTPointCloud2021,Mikuni:2021pou} have been studied, but no significant performance improvement over ParticleNet has been reported. Lately, researches have been focused more on incorporating inductive biases from physics principles in the architecture design, such as the usage of the Lund jet plane \cite{Dreyer:2018nbf,Dreyer:2020brq,Dreyer:2021hhr,Dreyer:2022yom}, the Lorentz group symmetry \cite{bogatskiy20a,Gong:2022lye}, and the rotational symmetry \cite{Shimmin:2021pkm,Dillon:2021gag}.

Deep-learning-based jet tagging algorithms have been widely adopted in real-world data analysis at the LHC. For example, the CMS Collaboration develops the DeepAK8 \cite{CMS:2020poo} algorithm to tag jets arising from the top quark or the Higgs, $W$, or $Z$ boson, using a 1D CNN following the ResNet \cite{heDeepResidualLearning2015} architecture, and a significant increase in the discovery potential for new heavy particles has been achieved \cite{CMS:2021beq,CMS:2022lqh}. Moreover, using ParticleNet, CMS achieves the first observation of $Z$ boson decay to a pair of charm quarks at a hadron collider and obtains the most stringent constraint on \hcc decay \cite{CMS:2022psv}. ParticleNet is also used by CMS to probe the quartic interaction between the Higgs and vector bosons, indirectly confirming its existence for the first time \cite{CMS:2022nmn}. Clearly, advances in jet tagging play a vital role in accelerating our understanding of elementary particles, the fundamental building blocks of nature.

\textbf{Jet tagging datasets.}
A number of datasets have been published so far to study jet tagging:
\begin{itemize}
    \vspace{-.2cm}
    \item \textbf{Top quark tagging dataset} \cite{kasieczkaTopQuarkTagging2019} proposed in \citet{Kasieczka:2019dbj}, consisting of 2\,M jets in 2 types (\tbqq and \qgj) and providing only the kinematic information.
    \vspace{-.2cm}
    \item \textbf{Quark-gluon tagging dataset} \cite{komiskePythia8QuarkGluon2019} proposed in \citet{Komiske:2018cqr}, consisting of 2\,M jets in 2 types (quark and gluon), and providing both the kinematic and particle identification information.
    \vspace{-.2cm}
    \item \textbf{Higgs boson tagging dataset} \cite{duarteSampleJetTrack2019,chenFAIRAIreadyHiggs2021}, containing 3.9\,M \hbb jets and 1.9\,M \qgj jets, with all three categories of information.  
    \vspace{-.2cm}
    \item \textbf{JetNet dataset} \cite{kansal_raghav_2021_5502543} proposed in \citet{kansalParticleCloudGeneration2021}, containing ${\approx}500\,\mathrm{k}$ jets in 3 types: gluon, light quark, and top quark, and providing only the kinematic information.
    \vspace{-.2cm}
    \item \textbf{A multiclass dataset} \cite{pierini_maurizio_2020_3602260} proposed in~\citet{Moreno:2019bmu}, with 880\,k jets in 5 classes: light quark, gluon, $W$ boson, $Z$ boson and top quark and providing only the kinematic information.
    \vspace{-.2cm}
\end{itemize}
Compared with existing datasets, the \jetclass dataset is not only substantially larger in size, but also more inclusive in terms of the types of jets contained.

\textbf{Transformers.}
Recent years have witnessed the enormous success of Transformer models. Starting from natural language processing and then spreading to computer vision, the original Transformer \cite{vaswaniAttentionAllYou2017}, as well as its variants, e.g., BERT \cite{devlinBERTPretrainingDeep2019}, ViT \cite{dosovitskiy2021an} and Swin-Transformer \cite{Liu_2021_ICCV}, have refreshed the performance records in various tasks, demonstrating the power of Transformer as a universal architecture. 
Transformers, and the attention mechanism at its core, have proved to be powerful for fundamental scientific problems as well. For example, AlphaFold2~\cite{Jumper2021}, which reaches the state-of-the-art performance in protein structure prediction, employs the attention mechanism. In particular, adding a pair bias, derived from pairwise features, to the self attention helps improve the model explainability.

\section{Model Architecture}
\label{sec:arch}

\begin{figure*}[t]
\begin{center}
\centerline{\includegraphics[width=0.85\textwidth]{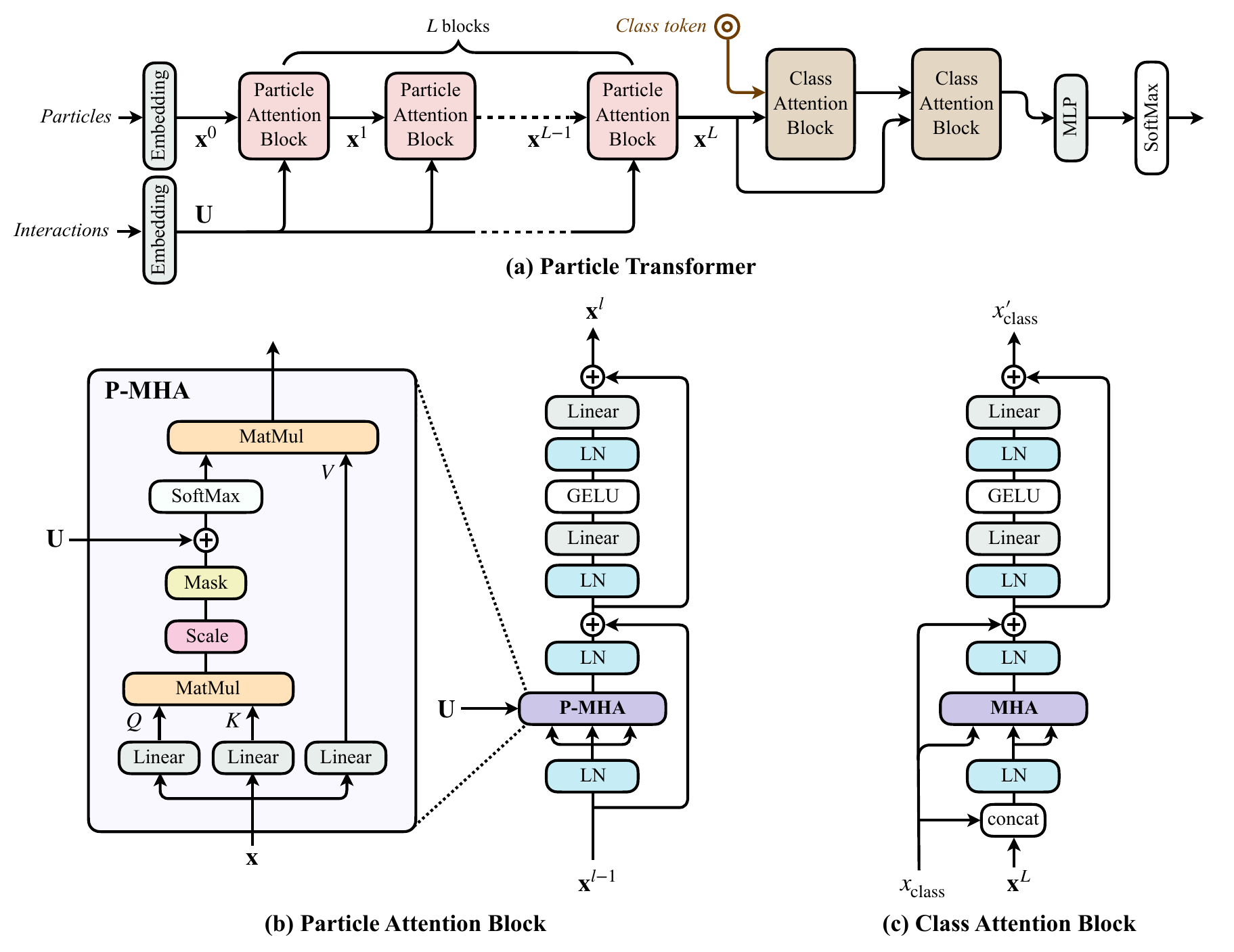}}
\vskip -0.2in
\caption{The architecture of (a) Particle Transformer (b) Particle Attention Block (c) Class Attention Block.}
\label{fig:arch}
\end{center}
\vskip -0.4in
\end{figure*}

Together with the \jetclass dataset, we propose the Particle Transformer (ParT) as a new baseline for jet tagging. An overview of the ParT architecture is presented in \cref{fig:arch}(a). For a jet with $N$ particles, ParT makes use of two sets of inputs: the \textit{particle} input includes a list of $C$ features for every particle and forms an array of a shape $(N, C)$; the \textit{interaction}\footnote{The term \textit{interaction} here refers to any feature involving a pair of particles, which may or may not be related to the physical forces between them.} input is a matrix of $C'$ features for every pair of particles, in a shape $(N, N, C')$. The particle and interaction inputs are each followed by an MLP to project them to a $d$- and $d'$-dimensional embedding, $\mathbf{x}^{0}\in\mathbb{R}^{N \times d}$ and $\mathbf{U}\in\mathbb{R}^{N \times N \times d'}$, respectively. Unlike Transformers for NLP and vision, we do not add any ad-hoc positional encodings, as the particles in a jet are permutation invariant. The spatial information (i.e., the flying direction of each particle) is directly included in the particle inputs. 
We feed the particle embedding $\mathbf{x}^{0}$ into a stack of $L$ particle attention blocks to produce new embeddings, $\mathbf{x}^{1}, ..., \mathbf{x}^{L}$ via multi-head self attention. The interaction matrix $\mathbf{U}$ is used to augment the scaled dot-product attention by adding it as a bias to the pre-softmax attention weights. The same $\mathbf{U}$ is used for all the particle attention blocks. After that, the last particle embedding $\mathbf{x}^{L}$ is fed into two class attention blocks, and a global class token $x_{\text{class}}$ is used to extract information for jet classification via attention to all the particles, following the CaiT approach \cite{Touvron_2021_ICCV}. The class token is passed to a single-layer MLP, followed by softmax, to produce the final classification scores. 

\textit{Remark.} ParT can also be viewed as a graph neural network on a fully-connected graph, in which each node corresponds to a particle, and the interactions are the edge features.

\textbf{Particle interaction features.} While the ParT architecture is designed to be able to process any kinds of pairwise interaction features, for this paper we only consider a specific scenario in which the interaction features are derived from the energy-momentum 4-vector, $p=(E, p_x, p_y, p_z)$, of each particle. This is the most general case for jet tagging, as the particle 4-vectors are available in every jet tagging task. Specifically, for a pair of particles $a$, $b$ with 4-vectors $p_{a}$, $p_{b}$, we calculate the following 4 features:
\begin{align}
\begin{split}
    \Delta &= \sqrt{(y_a - y_b)^2 + (\phi_a - \phi_b)^2}, \\
    k_{\text{T}} &= \min(p_{\text{T},a}, p_{\text{T},b}) \Delta, \\
    z &= \min(p_{\text{T},a}, p_{\text{T},b}) / (p_{\text{T},a} + p_{\text{T},b}), \\
    m^2 &= (E_a+E_b)^2 - \|\mathbf{p}_{a}+\mathbf{p}_{b}\|^2,
\label{eq:interaction}
\end{split}
\end{align}
where $y_i$ is the rapidity, $\phi_i$ is the azimuthal angle, $p_{\text{T},i} = (p_{x, i}^2+p_{y, i}^2)^{1/2}$ is the transverse momentum, and $\mathbf{p}_i=(p_{x,i}, p_{y,i}, p_{z,i})$ is the momentum 3-vector and $\|\cdot\|$ is the norm, for $i=a$, $b$. Since these variables typically have a long-tail distribution, we take the logarithm and use $(\ln \Delta, \ln k_{\text{T}}, \ln z, \ln m^2)$ as the interaction features for each particle pair. The choice of this set of features is  motivated by \citet{Dreyer:2020brq}.

\textbf{Particle attention block.} A key component of ParT is the particle attention block. As illustrated in \cref{fig:arch}(b), the particle attention block consists of two stages. The first stage includes a multi-head attention (MHA) module with a LayerNorm (LN) layer both before and afterwards. The second stage is a 2-layer MLP, with an LN before each linear layer and GELU nonlinearity in between. Residual connections are added after each stage. The overall block structure is based on NormFormer \cite{shleifer2021normformer}, however, we replace the standard MHA with P-MHA, an augmented version that can also exploit the pairwise particle interactions directly. The P-MHA is computed as 
\begin{equation}
    \text{P-MHA}(Q, K, V) = \text{SoftMax}(QK^T / \sqrt{d_k} + \mathbf{U}) V,
\end{equation}
where $Q$, $K$ and $V$ are linear projections of the particle embedding $\mathbf{x}^{l}$. Essentially, we add the interaction matrix $\mathbf{U}$ to the pre-softmax attention weights. This allows P-MHA to incorporate particle interaction features designed from physics principles and modify the dot-product attention weights, thus increasing the expressiveness of the attention mechanism.

\textbf{Class attention block.} As illustrated in \cref{fig:arch}(c), the class attention block has a similar structure as the particle attention block. However, unlike in the particle attention block where we compute the self attention between particles, here we compute the attention between a global class token $x_{\text{class}}$ and all the particles using the standard MHA. Specifically, the inputs to the MHA are 
\begin{align}
\begin{split}
    Q &= W_{q} x_{\text{class}} + b_{q}, \\
    K &= W_{k} \textbf{z} + b_{k}, \\
    V &= W_{v} \textbf{z} + b_{v},
\end{split}    
\end{align}
where $\textbf{z}=[x_{\text{class}}, \mathbf{x}^L]$ is the concatenation of the class token and the particle embedding after the last particle attention block, $\mathbf{x}^L$. 

\textbf{Implementation.} We implement the ParT model in PyTorch \cite{NEURIPS2019_9015}. Specifically, the P-MHA is implemented using the PyTorch's \texttt{MultiheadAttention} by providing the interaction matrix $\mathbf{U}$ as the \texttt{attn\_mask} input. The baseline ParT model has a total of $L=8$ particle attention blocks and 2 class attention blocks. It uses a particle embedding of a dimension $d=128$, encoded from the input particle features using a 3-layer MLP with (128, 512, 128) nodes each layer with GELU nonlinearity, and LN is used in between for normalization. The interaction input features are encoded using a 4-layer pointwise 1D convolution with (64, 64, 64, 8) channels with GELU nonlinearity and batch normalization in between to yield a $d'=8$ dimensional interaction matrix. The P-MHA (MHA) in the particle (class) attention blocks all have $d'=8$ heads, with a query dimension of 16 for each head, and an expansion factor of 4 for the MLP. We use a dropout of 0.1 for all particle attention blocks, and no dropout for the class attention block.
The choice of hyperparameters provides a reasonable baseline but is not extensively optimized.

\section{Experiments}
\label{sec:experiments}

We conduct experiments on the new \jetclass dataset and show the results in \cref{sec:exp-jet-class}. The pre-trained ParT models are also applied to two existing datasets with fine-tuning, and the performance is compared to previous state-of-the-arts in \cref{sec:exp-finetune}.

\subsection{Experiments on \textsc{JetClass} Dataset}
\label{sec:exp-jet-class}

\begin{table*}[tb]
\caption{Jet tagging performance on the \jetclass dataset. ParT is compared to PFN \cite{Komiske:2018cqr}, P-CNN \cite{CMS:2020poo} and the state-of-the-art ParticleNet \cite{Qu:2019gqs}. For all the metrics, a higher value indicates better performance. The ParT architecture using plain MHAs instead of P-MHAs, labelled as ParT (plain), is also shown for comparison. }
\label{tab:jet-class}
\begin{center}
\vskip -0.15in
\resizebox*{1\textwidth}{!}{
\begin{tabular}{lccccccccccc}
\toprule
                & \multicolumn{2}{c}{All classes}   & \hbb      & \hcc      & \hgg      & \hqqqq    & \hlvqq    & \tbqq     & \tblv     & \wqq      & \zqq      \\
                            & Accuracy  & AUC       &\rej{50\%} &\rej{50\%} &\rej{50\%} &\rej{50\%} &\rej{99\%} &\rej{50\%}&\rej{99.5\%}&\rej{50\%} &\rej{50\%} \\
\midrule
PFN                         & 0.772     & 0.9714    & 2924      & 841       & 75        & 198       & 265       & 797       & 721       & 189       & 159       \\
P-CNN                       & 0.809     & 0.9789    & 4890      & 1276      & 88        & 474       & 947       & 2907      & 2304      & 241       & 204       \\
ParticleNet                 & 0.844     & 0.9849    & 7634      & 2475      & 104       & 954       & 3339      & 10526     & 11173     & 347       & 283       \\
\bf ParT                    &\bf 0.861  &\bf 0.9877 &\bf 10638  &\bf 4149   &\bf 123    &\bf 1864   &\bf 5479   &\bf 32787  &\bf 15873  &\bf 543    &\bf 402    \\
\midrule
ParT (plain)                & 0.849     & 0.9859    & 9569      & 2911      & 112       & 1185      & 3868      & 17699     & 12987     & 384       & 311       \\
\bottomrule
\end{tabular}
}
\end{center}
\vskip -0.2in
\end{table*}

\textbf{Setup.} For experiments on the \jetclass dataset, we use the full set of particle features, including kinematics, particle identification, and trajectory displacement, as inputs. The full list of 17 features for each particle is summarized in \cref{tab:inputs}. In addition, the 4 interaction features introduced in \cref{eq:interaction} are also used for the ParT model. The training is performed on the full training set of 100\,M jets. We employ the Lookahead optimizer \cite{NEURIPS2019_90fd4f88} with $k=6$ and $\alpha=0.5$ to minimize the cross-entropy loss, and the inner optimizer is RAdam \cite{Liu2020On} with $\beta_1=0.95$, $\beta_2=0.999$, and $\epsilon=10^{{-}5}$. A batch size of 512 and an initial learning rate (LR) of 0.001 are used. No weight decay is applied. We train for a total of 1\,M iterations, amounting to around 5 epochs over the full training set. The LR remains constant for the first 70\% of the iterations, and then decays exponentially, at an interval of every 20\,k iterations, down to 1\% of the initial value at the end of the training. Performance of the model is evaluated every 20\,k iterations on the validation set and a model checkpoint is saved. The checkpoint with the highest accuracy on the validation set is used to evaluate the final performance on the test set. 

\begin{table*}[hbt]
\caption{Particle input features used for jet tagging on the \jetclass, the top quark tagging (\textsc{Top}) and the quark gluon tagging (\textsc{QG}) datasets. For \textsc{QG}, we consider two scenarios: \textsc{QG}\sst{exp} is restricted to use only the 5-class experimentally realistic particle identification information, while \textsc{QG}\sst{full} uses the full set of particle identification information in the dataset and further distinguish between different types of charged hadrons and neutral hadrons.}
\label{tab:inputs}
\vskip 0.05in
\begin{center}
\begin{minipage}{1\textwidth}
\renewcommand{\footnotesize}{\fontsize{7.5pt}{11pt}\selectfont}
\resizebox*{1\textwidth}{!}{
\begin{tabular}{cllcccc}
\toprule
Category                    & Variable                           & Definition                                                                   & \jetclass  & \textsc{Top} & \textsc{QG}\sst{exp} & \textsc{QG}\sst{full}  \\
\midrule                 
\multirow{7}{*}{Kinematics} & $\Delta \eta$                      & difference in pseudorapidity $\eta$ between the particle and the jet axis    & \checkmark & \checkmark   & \checkmark           & \checkmark             \\
                            & $\Delta \phi$                      & difference in azimuthal angle $\phi$ between the particle and the jet axis   & \checkmark & \checkmark   & \checkmark           & \checkmark             \\
                            & $\log \pt$                         & logarithm of the particle's transverse momentum \pt                          & \checkmark & \checkmark   & \checkmark           & \checkmark             \\
                            & $\log E$                           & logarithm of the particle's energy                                           & \checkmark & \checkmark   & \checkmark           & \checkmark             \\
                            & $\log\frac{\pt}{\pt(\text{jet})}$  & logarithm of the particle's \pt relative to the jet \pt                      & \checkmark & \checkmark   & \checkmark           & \checkmark             \\
                            & $\log \frac{E}{E(\text{jet})}$     & logarithm of the particle's energy relative to the jet energy                & \checkmark & \checkmark   & \checkmark           & \checkmark             \\
                            & $\Delta R$  & angular separation between the particle and the jet axis ($\sqrt{(\Delta\eta)^2 + (\Delta\phi)^2}$) & \checkmark & \checkmark   & \checkmark           & \checkmark             \\
\midrule
\multirow{6}{*}{\begin{tabular}{c}Particle\\identification\end{tabular}} & \texttt{charge}  & electric charge of the particle                   & \checkmark & ---          & \checkmark           & \checkmark             \\
                            & \texttt{Electron}                & if the particle is an electron (\texttt{|pid|==11})                            & \checkmark & ---          & \checkmark           & \checkmark             \\
                            & \texttt{Muon}                    & if the particle is an muon (\texttt{|pid|==13})                                & \checkmark & ---          & \checkmark           & \checkmark             \\
                            & \texttt{Photon}                  & if the particle is an photon (\texttt{pid==22})                                & \checkmark & ---          & \checkmark           & \checkmark             \\
                            & \texttt{CH}                      & if the particle is an charged hadron (\texttt{|pid|==211 or 321 or 2212})      & \checkmark & ---          & \checkmark           & \checkmark\footnote{\texttt{(|pid|==211) + (|pid|==321)*0.5 + (|pid|==2212)*0.2}}              \\
                            & \texttt{NH}                      & if the particle is an neutral hadron (\texttt{|pid|==130 or 2112 or 0})        & \checkmark & ---          & \checkmark           & \checkmark\footnote{\texttt{(|pid|==130) + (|pid|==2112)*0.2}.}            \\
\midrule
\multirow{4}{*}{\begin{tabular}{c}Trajectory\\displacement\end{tabular}} & $\tanh d_0$ & hyperbolic tangent of the transverse impact parameter value & \checkmark & ---     & ---                  & ---                    \\
                            & $\tanh d_z$                        & hyperbolic tangent of the longitudinal impact parameter value                & \checkmark & ---          & ---                  & ---                    \\
                            & $\sigma_{d_0}$                     & error of the measured transverse impact parameter                            & \checkmark & ---          & ---                  & ---                    \\
                            & $\sigma_{d_z}$                     & error of the measured longitudinal impact parameter                          & \checkmark & ---          & ---                  & ---                    \\
\bottomrule
\end{tabular}
}
\end{minipage}
\end{center}
\vskip -0.2in
\end{table*}

\textbf{Baselines.} We compare the performance of ParT with 3 baseline models: the PFN \cite{Komiske:2018cqr} architecture based on Deep Sets \cite{NIPS2017_f22e4747}, the P-CNN architecture used by the DeepAK8 algorithm of the CMS experiment \cite{CMS:2020poo}, and the state-of-the-art ParticleNet architecture \cite{Qu:2019gqs} adapted from DGCNN \cite{dgcnn}. All the models are trained end-to-end on the \jetclass dataset for the same number of effective epochs for a direct comparison. For ParticleNet, we directly use the existing PyTorch implementation. For PFN and P-CNN, we re-implement them in PyTorch and verify that the published results are reproduced. The optimizer and LR schedule remain the same as in the training of ParT. The (batch size, LR) combination is re-optimized and chosen to be (512, 0.01) for ParticleNet and (4096, 0.02) for PFN and P-CNN. 

\textbf{Results.} Performance on the \jetclass dataset is evaluated using the metrics described in \cref{sec:dataset}, and the results are summarized in \cref{tab:jet-class}. The proposed ParT architecture achieves the best performance on every metric, and outperforms the existing state-of-the-art, ParticleNet, by a large margin. The overall accuracy is increased by 1.7\% compared to ParticleNet. Moreover, for the physics-oriented metric, the background rejection, ParT improves over ParticleNet by a factor of 3 for \tbqq, a factor of 2 for \hqqqq, and about 70\% for \hcc. It is also clear that, the earlier PFN and P-CNN models lag substantially behind ParticleNet and ParT on this large dataset, amounting to up to an order of magnitude difference in background rejection. The large improvement of ParT is likely to lead to a significant jump in the discovery potential for related physics searches at the LHC. 

Another observation is that there is a large variation in tagging performance between signals of different types. The best separation against the background \qgj jets is achieved for \tblv and \hlvqq signals -- with the powerful ParT model, these two can be selected almost perfectly, i.e., at an efficiency of more than 99\% with nearly no contamination from background jets. This opens up new territory for jet tagging at the LHC, as these types of jets have not been exploited for tagging so far.

\textbf{Effectiveness of P-MHA.} To quantify the effectiveness of the P-MHA introduced in ParT, we carry out an ablation study by replacing the P-MHA with a standard MHA, the resulting architecture is then a plain Transformer and therefore denoted as ParT (plain). We train ParT (plain) with the same procedure as the full ParT and the performance is shown in \cref{tab:jet-class}. A drop of 1.2\% in accuracy is observed compared to the full ParT, and the background rejection is reduced by 20--30\% for most signals. Note that, replacing P-MHA with plain MHA implies that the particle interaction input is discarded completely, but this does not lead to any reduction of information content, as the interaction features defined in \cref{eq:interaction} are derived purely from the energy-momentum 4-vectors, which are already used as particle features via the 7 kinematic variables presented in \cref{tab:inputs}. Therefore, the improvement of ParT over a plain Transformer indeed arise from an efficient exploitation of the particle kinematic information using the P-MHA. 

{\vskip 0.15in}

\begin{table*}[tb]
\caption{Impacts of the training dataset size. Entries in bold correspond to the training using the full 100\,M training dataset.}
\label{tab:perf-vs-size}
\begin{center}
\vskip -0.15in
\resizebox*{1\textwidth}{!}{
\begin{tabular}{lccccccccccc}
\toprule
                            & \multicolumn{2}{c}{All classes}   & \hbb      & \hcc      & \hgg      & \hqqqq    & \hlvqq    & \tbqq     & \tblv     & \wqq      & \zqq      \\
                                        & Accuracy  & AUC       &\rej{50\%} &\rej{50\%} &\rej{50\%} &\rej{50\%} &\rej{99\%} &\rej{50\%}&\rej{99.5\%}&\rej{50\%} &\rej{50\%} \\
\midrule
ParticleNet (2\,M)            & 0.828     & 0.9820    & 5540      & 1681      & 90        & 662       & 1654      & 4049      & 4673      & 260       & 215       \\
ParticleNet (10\,M)           & 0.837     & 0.9837    & 5848      & 2070      & 96        & 770       & 2350      & 5495      & 6803      & 307       & 253       \\
\bf ParticleNet (100\,M)      & 0.844     & 0.9849    & 7634      & 2475      & 104       & 954       & 3339      & 10526     & 11173     & 347       & 283       \\
\midrule
ParT (2\,M)                   & 0.836     & 0.9834    & 5587      & 1982      & 93        & 761       & 1609      & 6061      & 4474      & 307       & 236       \\
ParT (10\,M)                  & 0.850     & 0.9860    & 8734      & 3040      & 110       & 1274      & 3257      & 12579     & 8969      & 431       & 324       \\
\bf ParT (100\,M)             & 0.861     & 0.9877    & 10638     & 4149      & 123       & 1864      & 5479      & 32787     & 15873     & 543       & 402       \\
\bottomrule
\end{tabular}
}
\end{center}
\vskip -0.2in
\end{table*}

\textbf{Impacts of the training dataset size.} To evaluate the impacts of the training dataset size on the jet tagging performance, we perform additional trainings using only 2\% and 10\% of the \jetclass dataset. For the former, the training is performed for only 100\,k iterations, as it is already converged by then. For the latter, the training still lasts for 1\,M iterations, although very little gain is observed compared to the training with only 100\,k iterations. No overfitting is found in either case. The results are summarized in \cref{tab:perf-vs-size}. For the ParticleNet model, a drop of 0.7\% in accuracy is observed when the training dataset size is reduced to 10\,M, and the drop in accuracy increases to 1.6\% when only 2\,M jets are used in the training. For the ParT model, the impact is even larger, the degradation in accuracy becomes 1.1\% and 2.5\% when the training dataset is reduced to 10\% and 2\%, respectively.

\textbf{Model complexity.} \cref{tab:cost} compares the model complexity of ParT with the baselines. While the number of trainable parameters is increased by more than $5\times$ compared to ParticleNet, the number of floating point operations (FLOPs) is actually 40\% lower. We also observe that the FLOPs of ParT are 30\% higher than ParT (plain), which mostly comes from the encoding of the pairwise features, because the computational cost there scales quadratically with the number of particles in a jet.

\begin{table}[bt]
\caption{Number of trainable parameters and FLOPs.}
\label{tab:cost}
\vskip 0.05in
\begin{center}
\resizebox*{0.7\columnwidth}{!}{
\begin{tabular}{lccc}
\toprule
                            & Accuracy  & \# params & FLOPs     \\
\midrule
PFN                         & 0.772     & 86.1\,k   & 4.62\,M   \\
P-CNN                       & 0.809     & 354\,k    & 15.5\,M   \\
ParticleNet                 & 0.844     & 370\,k    & 540\,M    \\
\bf ParT                    &\bf 0.861  & 2.14\,M   & 340\,M    \\
\midrule
ParT (plain)                & 0.849     & 2.13\,M   & 260\,M    \\
\bottomrule
\end{tabular}
}
\end{center}
\vskip -0.2in
\end{table}

\subsection{Fine-Tuning for Other Datasets}
\label{sec:exp-finetune}

\textbf{Top quark tagging dataset.} The top quark tagging benchmark \cite{Kasieczka:2019dbj} provides a dataset of 2\,M (1.2/0.4/0.4\,M for train/validation/test) jets in two classes, \tbqq (signal) and \qgj (background). Only kinematic features, i.e., the energy-momentum 4-vectors, are provided. Therefore, we pre-train a ParT model on the \jetclass dataset using only the kinematic features, and then fine-tune it on the top quark tagging dataset. The particle input features are the 7 kinematic features listed in \cref{tab:inputs}, the same as used by ParticleNet. The \jetclass pre-training follows the same setup as described in \cref{sec:exp-jet-class}. For the fine-tuning, we replace the last MLP with a new randomly-initialized MLP with 2 output nodes, and then fine-tune all the weights on the top tagging dataset for 20 epochs. A smaller LR of 0.0001 is used for the pre-trained weights, while a larger LR of 0.005 is used to update the randomly-initialized weights of the MLP. The LR remains constant across the full training, with a weight decay of 0.01. We run a total of 9 experiments, starting from the same pre-trained model but different random initializations of the replaced MLP, and report the performance of the model with median accuracy and the spread across the 9 trainings, following the procedure used by ParticleNet. For comparison, we also train ParT from scratch on this dataset for 20 epochs, using a start LR of 0.001, a schedule that decays the LR to 1\% in the last 30\% of the epochs, and a weight decay of 0.01. Both results are presented in \cref{tab:top-tagging}. The pre-trained ParT achieves a significant improvement over the existing baselines, increasing \rej{30\%} by 70\% compared to ParticleNet, and by 26\% compared to the best-performing model on this dataset, LorentzNet. On the other hand, the ParT model trained from scratch only reaches similar performance as ParticleNet. We also investigate a similar pre-training and fine-tuning procedure using the ParticleNet model, but only a small improvement is observed compared to the training from scratch, due to the limited capacity of the ParticleNet model.

\begin{table}[tb]
\caption{Comparison between ParT and existing models on the top quark tagging dataset. ParT refers to the model trained from scratch on this dataset. ParticleNet-f.t. and ParT-f.t. denote the corresponding models pre-trained on \jetclass and fine-tuned on this dataset. Results for other models are quoted from their published results: P-CNN and ParticleNet \cite{Qu:2019gqs}, PFN \cite{Komiske:2018cqr}, JEDI-net \cite{Moreno:2019bmu}, PCT \cite{Mikuni:2021pou}, LGN \cite{bogatskiy20a}, rPCN \cite{Shimmin:2021pkm}, and LorentzNet \cite{Gong:2022lye}.}
\label{tab:top-tagging}
\begin{center}
\vskip -0.15in
\resizebox*{1\columnwidth}{!}{
\begin{tabular}{lcccc}
\toprule
                        & Accuracy   & AUC       & \rej{50\%}    & \rej{30\%}        \\
\midrule        
P-CNN                   & 0.930      & 0.9803    & $201\pm4$     & $759\pm24$        \\
PFN                     & ---        & 0.9819    & $247\pm3$     & $888\pm17$        \\
ParticleNet             & 0.940      & 0.9858    & $397\pm7$     & $1615\pm93$       \\
JEDI-net (w/ $\sum O$)  & 0.930      & 0.9807    & ---           & 774.6             \\
PCT                     & 0.940      & 0.9855    & $392\pm7$     & $1533\pm101$      \\
LGN                     & 0.929      & 0.964     & ---           & $435\pm95$        \\
rPCN                    & ---        & 0.9845    & $364\pm9$     & $1642\pm93$       \\
LorentzNet              & 0.942      & 0.9868    & $498\pm18$    & $2195\pm173$       \\
ParT                    & 0.940      & 0.9858    & $413\pm16$    & $1602\pm81$       \\
ParticleNet-f.t.        & 0.942      & 0.9866    & $487\pm9$     & $1771\pm80$       \\
\bf ParT-f.t.           & \bf 0.944  &\bf 0.9877 & $\bf691\pm15$ & $\bf2766\pm130$   \\
\bottomrule
\end{tabular}
}
\end{center}
\vskip -0.25in
\end{table}

\textbf{Quark-gluon tagging dataset.} We also benchmark ParT on the quark-gluon tagging dataset \cite{komiskePythia8QuarkGluon2019} proposed in \citet{Komiske:2018cqr}, the target of which is to separate jets initiated by quarks (signal) from those by gluons (background). This dataset also consists of 2\,M jets, with a recommended train/validation/test splitting of 1.6/0.2/0.2\,M. It provides not only the kinematic features, but also particle identification information. We consider two scenarios in the usage of the particle identification information. In the ``exp'' scenario, we restrict the information to only 5 classes and do not attempt to separate electrically charged (and neural) hadrons of different types, which is the procedure adopted by ParticleNet, and also prescribed by the \jetclass dataset. In the ``full'' scenario, we consider all particle types and further distinguish electrically charged (and neural) hadrons into more types, such as pions, kaons, and protons. 
We perform the pre-training on \jetclass using only kinematic and particle identification inputs under the ``exp'' scenario. For the fine-tuning, we then carry out experiments in both scenarios. The construction of the input features is described in \cref{tab:inputs}. 
The pre-training and fine-tuning setup is the same as in the top quark tagging benchmark, and the fine-tuning also lasts for 20 epochs. Results are summarized in \cref{tab:quark-gluon}. The pre-trained ParT achieves the best performance and improves existing baselines by a large margin in both scenarios.

\begin{table}[tb]
\caption{Comparison between ParT and existing models on the quark-gluon tagging dataset. ParT refers to the model trained from scratch on this dataset. ParticleNet-f.t. and ParT-f.t. denote the corresponding models pre-trained on \jetclass and fine-tuned on this dataset. Results for other models are quoted from their published results: P-CNN and ParticleNet \cite{Qu:2019gqs}, PFN \cite{Komiske:2018cqr}, ABCNet \cite{Mikuni:2020wpr}, PCT \cite{Mikuni:2021pou}, rPCN \cite{Shimmin:2021pkm}, and LorentzNet \cite{Gong:2022lye}. The subscript ``exp'' and ``full'' distinguish models using partial or full particle identification information.}
\label{tab:quark-gluon}
\begin{center}
\vskip -0.15in
\resizebox*{1\columnwidth}{!}{
\begin{tabular}{lcccc}
\toprule
                            & Accuracy   & AUC       & \rej{50\%}    & \rej{30\%}         \\
\midrule
P-CNN\sst{exp}              & 0.827      & 0.9002    & 34.7          & 91.0               \\
PFN\sst{exp}                & ---        & 0.9005    & $34.7\pm0.4$  & ---                \\
ParticleNet\sst{exp}        & 0.840      & 0.9116    & $39.8\pm0.2$  & $98.6\pm1.3$       \\
rPCN\sst{exp}               & ---        & 0.9081    & $38.6\pm0.5$  & ---                \\
ParT\sst{exp}               & 0.840      & 0.9121    & $41.3\pm0.3$  & $101.2\pm1.1$      \\
ParticleNet-f.t.\sst{exp}   & 0.839      & 0.9115    & $40.1\pm0.2$  & $100.3\pm1.0$      \\
\bf ParT-f.t.\sst{exp}      &\bf 0.843   &\bf 0.9151 &$\bf42.4\pm0.2$&$\bf107.9\pm0.5$    \\

\midrule
PFN\sst{full}               & ---        & 0.9052    & $37.4\pm0.7$  & ---                \\
ABCNet\sst{full}            & 0.840      & 0.9126    & $42.6\pm0.4$  & $118.4\pm1.5$      \\
PCT\sst{full}               & 0.841      & 0.9140    & $43.2\pm0.7$  & $118.0\pm2.2$      \\
LorentzNet\sst{full}        & 0.844      & 0.9156    & $42.4\pm0.4$  & $110.2\pm1.3$      \\
ParT\sst{full}              & 0.849      & 0.9203    & $47.9\pm0.5$  & $129.5\pm0.9$      \\
\bf ParT-f.t.\sst{full}     &\bf 0.852   &\bf 0.9230 &$\bf50.6\pm0.2$&$\bf138.7\pm1.3$    \\
\bottomrule
\end{tabular}
}
\end{center}
\vskip -0.25in
\end{table}

\section{Discussion and Conclusion}
\label{sec:conclusion}

Large-scale datasets have always been a catalyst for new breakthroughs in deep learning. In this work, we present \jetclass, a new large-scale open dataset to advance deep learning research in particle physics. The dataset consists of 100\,M simulated jets, about two orders of magnitude larger than existing public jet datasets, and covers a broad spectrum of 10 classes of jets in total, including several novel types that have not been studied with deep learning so far. While we focus on investigating a classification task, i.e., jet tagging, with this dataset, we highlight that this dataset can serve as the basis for many important deep learning researches in particle physics, e.g., unsupervised or self-supervised training techniques for particle physics (e.g., \citet{Dillon:2021gag}), generative models for high-fidelity fast simulation of particle collisions (e.g., \citet{kansalParticleCloudGeneration2021}), regression models to predict jet energy and momentum with higher precision (e.g., \citet{CMS:2019uxx}), and more. We invite the community to explore and experiment with this dataset and extend the boundary of deep learning and particle physics even further.

With this large dataset, we introduce Particle Transformer (ParT), a new architecture that substantially improves jet tagging performance over previous state-of-the-art. We propose it as a new jet tagging baseline for future research to improve upon. The effectiveness of ParT arises mainly from the augmented self-attention, in which we incorporate physics-inspired pairwise interactions together with the machine-learned dot-product attention. This approach is likely to be effective for other tasks on similar datasets, such as point clouds or many-body systems, especially when prior knowledge is available to describe the interaction or the geometry. On the other hand, one limitation of using the full pairwise interaction matrix is the increase in computational time and memory consumption. Novel approaches for particle (point) embeddings and self-attentions that alleviate the computational cost (e.g., \citet{Informer,Kitaev2020Reformer}) could be an interesting direction for future research.

\section*{Acknowledgements}

We are grateful to Loukas Gouskos, Qiang Li, and Alexandre De Moor for many helpful discussions. 
The work of C.~Li and S.~Qian is supported by National Natural Science Foundation of China under Grants No.~12061141002.

\bibliography{main} 
\bibliographystyle{icml2022}


\end{document}